\documentclass[aps,prl,superscriptaddress,reprint]{revtex4-1}
\usepackage{amsmath}    
\usepackage{graphicx}   
\usepackage[dvipdfx,cmyk]{xcolor}     
\usepackage[caption=false]{subfig}  
\usepackage[colorlinks,linkcolor=red,citecolor=brown,urlcolor=blue]{hyperref}
\usepackage{dsfont,bm}
\usepackage{color}
\usepackage{placeins}

\usepackage{amsmath} 
\usepackage{amsthm} 
\usepackage{amssymb}	
\usepackage{graphicx} 
\renewcommand{\v}[1]{\ensuremath{\boldsymbol{#1}}} 

\newcommand{\abs}[1]{\left| #1 \right|} 
\newcommand{\pd}[2]{\frac{\partial #1}{\partial #2}}

\newcommand{\ket}[1]{\left| #1 \right>} 
\newcommand{\bra}[1]{\left< #1 \right|} 
\let\baraccent=\= 
\renewcommand{\=}[1]{\stackrel{#1}{=}} 

\theoremstyle{definition}

\theoremstyle{remark}

\graphicspath{{Figures/}}

\begin{document}
\title{Tomography of photon-number resolving continuous-output detectors}

\author{Peter C. Humphreys}
\affiliation{Clarendon Laboratory, Department of Physics, University of Oxford, OX1 3PU, United Kingdom}
\author{Benjamin J. Metcalf}
\affiliation{Clarendon Laboratory, Department of Physics, University of Oxford, OX1 3PU, United Kingdom}
\author{Thomas Gerrits}
\affiliation{National Institute of Standards and Technology, Boulder, CO, 80305, USA}
\author{Thomas Hiemstra}
\affiliation{Clarendon Laboratory, Department of Physics, University of Oxford, OX1 3PU, United Kingdom}
\author{Adriana E. Lita}
\affiliation{National Institute of Standards and Technology, Boulder, CO, 80305, USA}
\author{Joshua Nunn}
\affiliation{Clarendon Laboratory, Department of Physics, University of Oxford, OX1 3PU, United Kingdom}
\author{Sae Woo Nam}
\affiliation{National Institute of Standards and Technology, Boulder, CO, 80305, USA}
\author{Animesh Datta}
\affiliation{Clarendon Laboratory, Department of Physics, University of Oxford, OX1 3PU, United Kingdom}
\author{W. Steven Kolthammer}
\affiliation{Clarendon Laboratory, Department of Physics, University of Oxford, OX1 3PU, United Kingdom}
\author{Ian A. Walmsley}
\affiliation{Clarendon Laboratory, Department of Physics, University of Oxford, OX1 3PU, United Kingdom}

\begin{abstract}
We report a comprehensive approach to analysing continuous-output photon detectors. We employ principal component analysis to maximise the information extracted, followed by a novel noise-tolerant parameterised approach to the tomography of PNRDs. We further propose a measure for rigorously quantifying a detector's photon-number-resolving capability. Our approach applies to all detectors with continuous-output signals. We illustrate our methods by applying them to experimental data obtained from a transition-edge sensor (TES) detector.
\end{abstract}
\maketitle

The continuing development of highly efficient photon detectors has significant impact across a broad range of fields, from quantum information~\cite{Calkins2013} to astronomy~\cite{Day2003} and biomedical imaging~\cite{art2006photon}. The physics underlying the operation of different photon detectors is rich and varied, but their outputs typically fall into two categories. Those such as photomultiplier tubes, avalanche photodiodes~\cite{Eisaman2011} and superconducting nanowires~\cite{Natarajan2012,Marsili2013a} are often based on avalanche phenomena and lead to discrete `click' outcomes, while others, such as transition-edge sensors~\cite{Lita2008}, kinetic-inductance detectors~\cite{Day2003} and superconducting tunnel junctions~\cite{Peacock1996} rely on smooth transitions leading to continuous `trace' outputs (avalanche photodiodes can also give continuous-valued outputs under appropriate conditions~\cite{Kardyna2008}). Some of these, including TES detectors, are highly sensitive single-photon detectors with quantum efficiencies of up to 98\%~\cite{Lita2008,Fukuda2011} and true photon-number sensitivity~\cite{Lita2008}. Others, such as microwave kinetic inductance detectors, allow unprecedented level of integration into large arrays~\cite{Day2003}. These advances over traditional discrete-output detectors will enable new applications in wide-ranging fields. 

With these novel applications and regimes of performance come additional challenges in detector characterisation. Unlike discrete-output detectors, many photon-number resolving detectors (PNRD) produce a complex time-varying signal from which the input state must be inferred. Efficiently extracting information from these signals is therefore necessary to realise the full capability of such detectors~\cite{Avella2011,Brida2012a,Levine2012}.

\begin{figure}[htbp]
\begin{center}
\includegraphics[width=8.0cm]{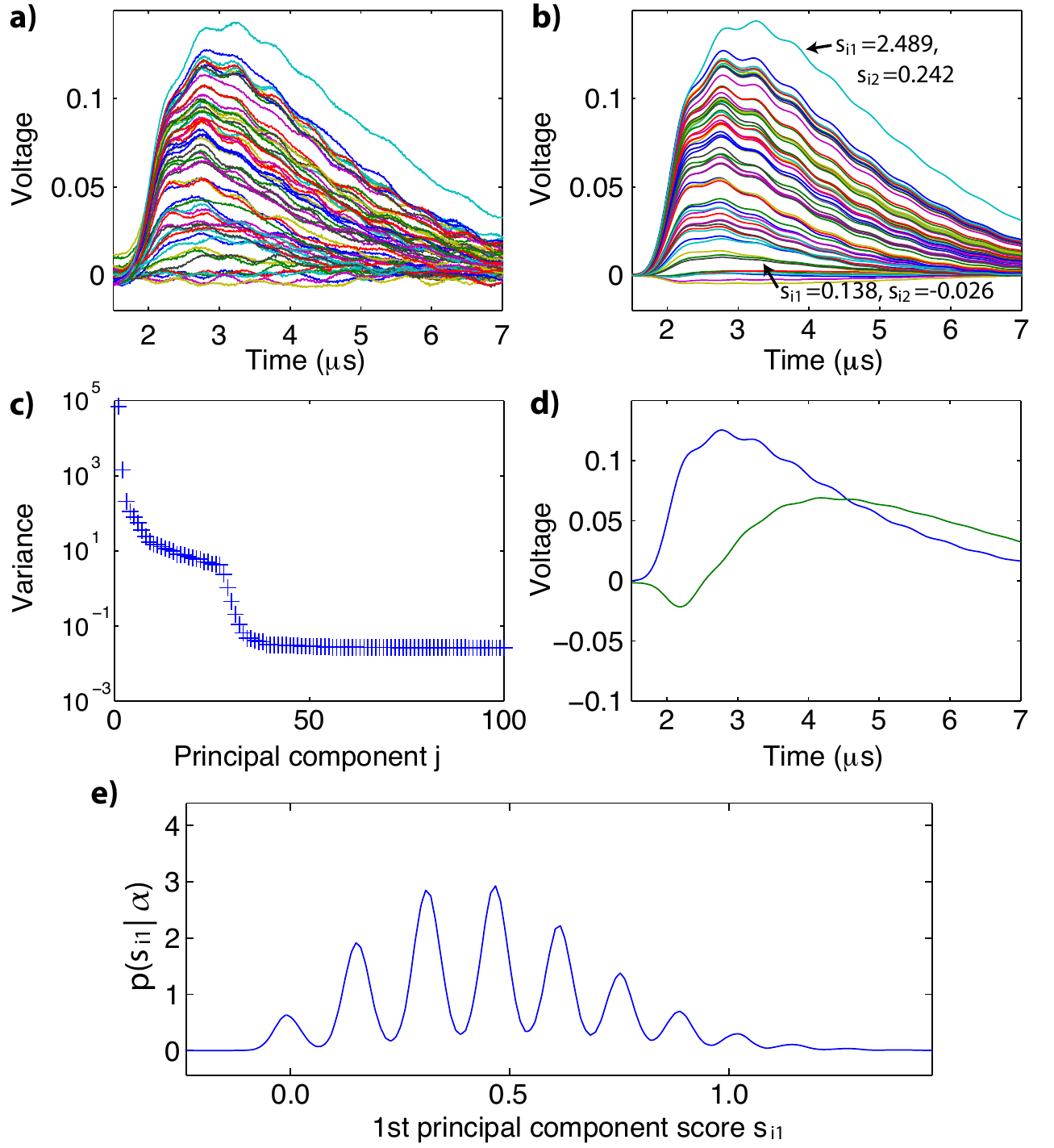}
\caption{a) Representative TES traces $v_i(t)$ from a data set of 180,000 total signals. b) Truncated representation of the same traces using only the first two principal components $w_1(t)$ and $w_2(t)$. c) Variance of the set of principal component scores $\{ s_{ij} \}$ as a function of the principal component number $j$. d) Principal components $w_1(t)$ (blue) and $w_2(t)$ (green). e) Probability density function $p(s_{i1} | \alpha)$ over the signal scores $s_{i1}$ for a coherent state input $\ket{\alpha}\!\!\bra{\alpha}$ with a mean of 3.1 photons per pulse. Note that these values can be negative as the mean signal is subtracted from each signal during the calculation of $s_{i1}$}
\label{fig:SVDtraces}
\end{center}
\end{figure}

The signal produced by a continuous-output detector is typically a time-dependent voltage with some dependence on photon number which may in general be nonlinear, as shown in Fig.~\ref{fig:SVDtraces}a. A set of such output signals $\v{V} = \{v_i(t)\}$, arising from a set of input states of the incident light beam, can be represented using a set of basis functions $\{w_j (t)\}$, such that
\begin{align}
v_i(t) = \sum^n_j s_{ij} w_j(t). \notag
\end{align}
In general, this implies that, in order to capture the full output of the detector, it is necessary to determine the weighting components $s_{ij}$ for all of the $n$ basis functions for each signal to be measured. For a truly continuous signal, $n$ is in principle infinite, but of course for any real experiment the upper limit to $n$ is set by the temporal and voltage resolution of the detector. However, this finite signal still spans a space of high dimension; in our work a signal consists of 1024 16-bit numbers. Directly analysing this signal is therefore impractical. This is particularly the case for detector tomography, necessary to rigorously characterise the relationship between input states and output signals~\cite{Lundeen2008,Zhang2012a}. Detector tomography requires a sufficiently small space of outputs that the probability of a given outcome can be estimated precisely from the measured data. For the full output space of our detector signal, we estimate the probability of the same trace occurring twice (to within the resolution of the analogue-to-digital converter) in a data set of $10^5$ traces to be on the order of $10^{-4}$, rendering tomography in this full space infeasible. This motivates the development of an approach to the characterization of continous-output detectors that enables accurate and precise signal analysis and detector tomography. 

Detector tomography has been previously carried out for continuous-output PNRDs with 5\% quantum efficiencies~\cite{Brida2012a}, in which the continuous-output problem was circumvented by `binning' the detector output based on the maximum amplitude of the signal. This approach does not make optimal use of the information available. Furthermore, as we will discuss, the numerical techniques for detector tomography used in the study are not effective in the high detection-efficiency regime, which is now accessible with TES detectors. Another recent work has explored algorithmic methods of interpreting the response of high detection-efficiency PNRDs based on cluster analysis~\cite{Levine2012}. Although this may prove useful for rapid characterisation of a detector, it is not a tomographic technique and is therefore unable to provide a rigorous characterisation of the detector response. 


\emph{Principal component analysis:} We first consider the problem of efficiently extracting information from a high-dimensional detector signal data set. We achieve this by employing a standard technique from multi-variate statistics, namely principal component analysis~\cite{Abdi2010}. For a given data set, this approach determines the optimal set of `principal component' basis functions $\{w_j (t)\}$ such that each successive basis function captures the maximum amount of information possible from the data set (as measured by the variance of the `principal component scores' $s_{ij}$), while maintaining orthogonality with the previous components. Crucially, this implies that if the principal component basis is truncated to compress the data, the maximum amount of the variance of the original data set will still be captured. In other words, the truncated principal component basis will provide the most faithful reconstruction of the data for a given number of components. 

In an actual experiment, the signals $v_i(t)$ and therefore the basis functions $w_j (t)$ are necessarily discretised due to the finite temporal resolution of the detector. In this case the set of signals $\v{V}$ can be expressed as a matrix. It can be shown that the problem of determining $\{w_j (t)\}$ for $\v{V}$ is equivalent to finding the eigenvectors of the matrix $\v{\tilde{V}}^T\v{\tilde{V}}$, where $\v{\tilde{V}}$ is the data set with the mean signal subtracted~\cite{Abdi2010}. These eigenvectors can be efficiently determined using singular value decomposition. Once $w_i(t)$ are known, $s_{ij}$ can be calculated from the detector signals $v_j(t)$ by $s_{ij} = \int v_j(t) w_i(t) \mathrm{d}t$. 

We applied principal component analysis to a data set of 180,000 TES traces, taken with a range of 300 different coherent state inputs with average photon numbers spanning from 0 to approximately 15 photons per pulse. In Fig.~\ref{fig:SVDtraces}a~\&~b, example TES traces from this data set are plotted both in their original form, and in a reduced form using only the first two principal components $w_1(t)$ and $w_2(t)$. As can be seen, with just these two components, most of the structure of the traces has been reproduced. This can be shown more formally by comparing the variance of $\{ s_{ij} \}$ for different principal component numbers $j$, as plotted in Fig.~\ref{fig:SVDtraces}c. The variance of $\{ s_{i1} \}$ is two orders of magnitude greater than $\{ s_{i2} \}$, and this trend continues, with the variance rapidly decreasing as a function of $j$.

Interestingly, as Fig.~\ref{fig:SVDtraces}d shows, $w_1(t)$ is very close to the mean shape of the TES traces. This would be expected theoretically in the small-signal limit, in which the TES trace height simply scales linearly with the photon number~\cite{Miller2001}. This confirms that projecting onto the mean trace shape, as used by \cite{Levine2012}, is a useful approach for distinguishing TES signals in the few-photon limit using only a single parameter. Beyond providing a justification for this choice of processing method, the higher order principal components that are revealed by our analysis can provide additional data with which to characterise the response of a detector, particularly for higher photon numbers. For example, $w_2(t)$ captures the increase in the pulse length with photon number due to an increase in thermal recovery time~\footnote{See Supplementary Information: Section I for further discussion on the second principal component scores for the coherent probe state data.}. However, since the dominant contribution to the data variance is from $w_1(t)$, particularly for the low photon numbers considered here, we choose to solely focus on this component for the remainder of our analysis.

%



\emph{Detector tomography:} We now seek to determine the correspondence between the reduced detector signals and the input number of photons by carrying out detector tomography~\cite{Lundeen2008}. The goal of detector tomography is to determine the positive-operator-valued measure (POVM) $\{ \pi(s) \}$ that fully characterises the detector response; this is parameterised by the outcome $s$ in the space of $s_{i1}$. Once the POVM is known, the probability density for detector outcome $s$, given input state $\rho$, is determined by the Born rule
\begin{equation}
p(s | {\rho}) = \mathrm{Tr}\left [\rho \, \pi(s) \right ] \label{eqn:born}.
\end{equation}
The standard approach to tomography consists of experimentally estimating the outcome probability densities $p(s | \rho_k)$ for a set of known probe basis states $\left \{ \rho_k \right \}$. Using these estimated probabilities, equation~(\ref{eqn:born}) can then, in principle, be inverted to find $\pi(s)$. 

The set of probe states $\left \{ \rho_k \right \}$ must provide a sufficient basis for the operator space of the POVM $\{ \pi(s) \}$; in other words, it must be tomographically complete. We satisfy this constraint by using a well established method~\cite{Lundeen2008} for tomography of PNRDs based on coherent state probes $\ket{\alpha}$. It is well known that coherent states form an over-complete basis for an optical mode. Coherent states are also straightforward to generate in the lab and are insensitive in form to experimental losses during preparation, making them ideal probe states. Additionally, as TES detectors are phase insensitive, their response depends only on the magnitude of the coherent state parameter $\alpha$, and not its phase. This significantly reduces the number of probe states needed to form a tomographically complete set of basis operators and removes the need for any phase reference in the experiment. 

A phase insensitive detector will have POVM elements diagonal in the photon-number basis; these can therefore be expressed as 
\begin{equation}
\pi(s) = \sum_{n=0}^{\infty} \theta_{n}(s) \ket{n}\!\!\bra{n}.	\label{eqn:fockStateBasis}
\end{equation}
Coherent-state probes are given in this basis by~\footnote{This assumes that the input state is a pure state, however this can easily be extended to mixed state inputs~\cite{Lundeen2008} (due to classical uncertainties in $\alpha$ for example).} 
\begin{equation}
\ket{\alpha} = \exp(-\abs{\alpha}^2/2)\sum_{n=0}^{\infty} \frac{\alpha^{n}}{\sqrt{n!}} \ket{n},\label{eqn:coherentState}
\end{equation}
Inserting equations~(\ref{eqn:fockStateBasis})~\&~(\ref{eqn:coherentState}) into the Born rule (equation~(\ref{eqn:born})), we find that the probability density for a given outcome is
\begin{equation}
p(s | \alpha)  = \sum_{n=0}^{\infty} \, F_{\alpha,n} \,\theta_{n}(s).  \label{eqn:prob}
\end{equation}
where $F_{\alpha,n} = \abs{\alpha}^{2n} \frac{\exp(-\abs{\alpha}^2)}{n!}$

Using the set of probability density functions $p(s | \alpha_k)$ associated with the input probe states $\{ \ket{\alpha_k} \}$, this relation can be numerically inverted to find the best solution for $\theta_{n}(s)$ consistent with the physicality constraints
\begin{equation}
\theta_{n}(s) \geq 0, \qquad \text{and}  \qquad \int  \theta_{n}(s) \, \mathrm{d}s \leq 1. \notag
\end{equation}


It is necessary to use a calibrated light source in order to produce coherent-state probes with known energies for detector tomography. Since we do not have access to a source calibrated to a radiometric standard, we built our own calibrated source by using a Newport 918D-IG-OD3R power meter, which provides a specified calibration accuracy of 2\% of absolute power and a linearity of better than 0.5\%. This power meter was used to calibrate a series of fixed attenuators to reduce the output from a pulsed laser to the single-photon level with a known mean-photon number per pulse~\footnote{See Supplementary Information: Section III for further details on our calibrated coherent state source.}.

We measured the detector response to a set of 300 different probe energies equally spaced between 0 and 15 photons per pulse. For each probe energy, we ran 49152 trials, and used the measured signals to estimate the probability density function for the outcomes in the space of $s_{i1}$~\footnote{The probability density functions were calculated using Gaussian-kernel density estimation~\cite{Botev2010}. This technique is better suited to this problem than using histograms, as it is not necessary to choose an arbitrary binning of the data. Instead, this approach directly gives continuous-valued estimates of the functions.}.  Fig.~\ref{fig:SVDtraces}e shows an example measured probability density function for a probe state with a mean of 3.1 photons per pulse.

It is well known that the problem of inverting equation~(\ref{eqn:born}) to obtain $\pi(s) $ is ill-conditioned~\cite{Lundeen2008}. We found that published methods of performing this numerical inversion based on constrained least squares techniques~\cite{Brida2012a} did not give satisfactory results~\footnote{See Supplementary Information: Section II for our model-free tomography results.}. This may be in part due to the reduced overlap between the POVM elements for different photon numbers as compared to previous studies because of our much higher system detection efficiency. This means that regularisation techniques designed to promote this overlap~\cite{Lundeen2008,Zhang2012a} do not work as effectively. 

We used insights from our collected data to develop a novel detector tomography routine that is effective for high quantum efficiencies. We adopted a model in which the detector response to photon number $n$ (in the space of $S_1$) is given by the sum of $n + 1$ Gaussians, with widths, heights, and positions as free variables. This Gaussian-mixture model~\cite{Bishop2006} is consistent with detectors for which several different sources of noise contribute to the response of the detector to a given photon number, leading to an overall Gaussian error as might be expected from the central limit theorem.  We employed a maximum likelihood routine~\footnote{See Supplementary Information: Section II for more details on our maximum likelihood detector tomography routine.} to find the parameterised POVM that was most consistent with the full coherent state tomography data set. 

\begin{figure}[htbp]
\begin{center}
\includegraphics[width=7.5cm]{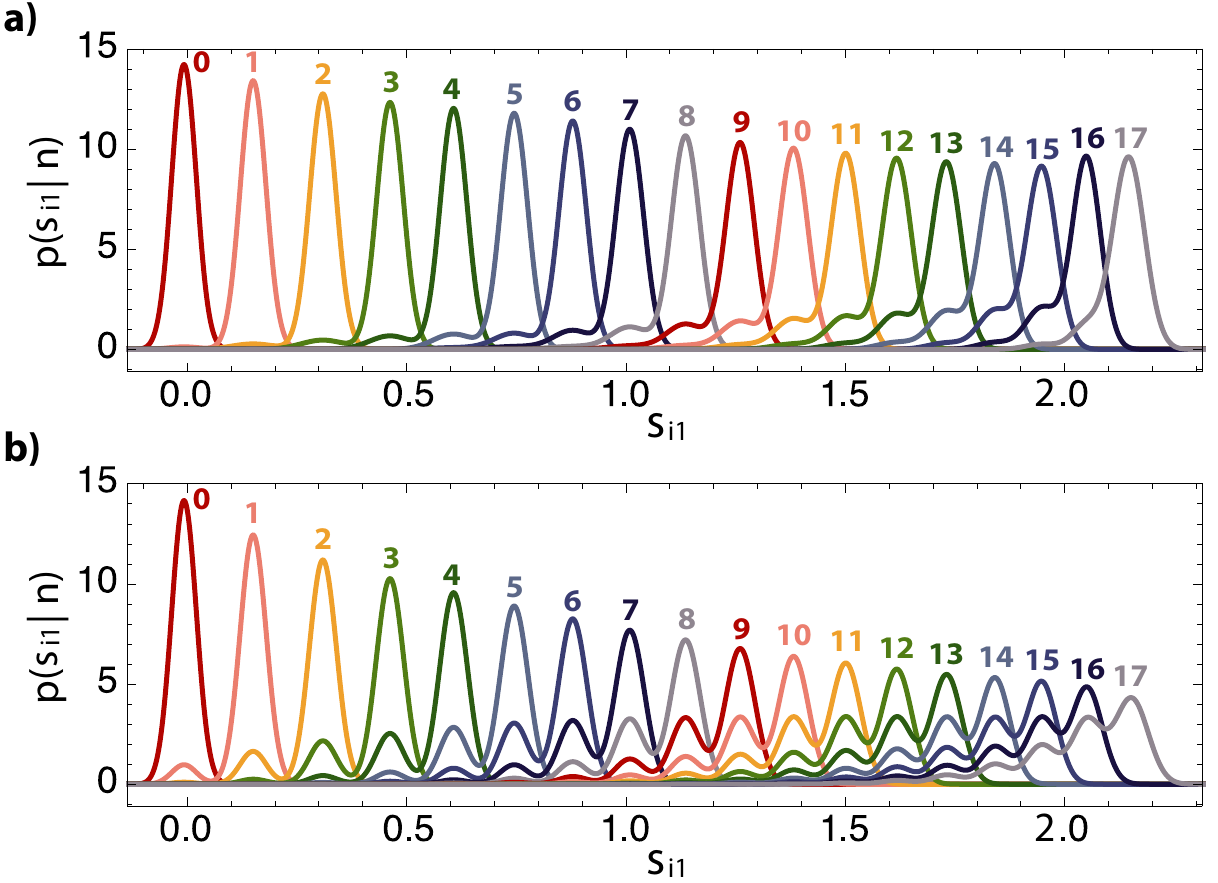}
\caption{a) Fock state POVM elements determined from our parameterised detector tomography routine. Note that these solutions are continuous functions in the space of $s_{i1}$, and have not been arbitrarily binned into different `photon-number' outcomes. b) Fock state POVM elements after incorporating the uncertainty in the probe state energies.}
\label{fig:EMsolutionAndCalib}
\end{center}
\end{figure}

The results of this inversion are shown in Fig.~\ref{fig:EMsolutionAndCalib}a. The efficacy of this model-based routine can be estimated by using the calculated POVM to reconstruct the original data set. The $L_1$ difference between this reconstruction and the original data (normalised by the $L_1$ norm of the original data set) is 0.054 as compared to 0.047 for the unphysical reconstruction given by a least-squares approach, showing that this model equally effectively captures the detector response while being significantly more robust to noise. The model-based approach also allows us to estimate the system detection efficiency from the tomography data, giving an efficiency of $0.98 \, ({+0.02}/{-0.08})$~\footnote{The system detection efficiency is defined as the efficiency with which a photon in the fiber connected to the detector is detected~\cite{Marsili2013a}. See Supplementary Information: Section IV for details on estimating the system detection efficiency.}. 

As a final step, it is necessary to incorporate the uncertainty in the coherent-state probe energies~\footnote{See Supplementary Information: Section III for additional details on our the calibration factor uncertainty analysis.} to give the POVM elements shown in Fig.~\ref{fig:EMsolutionAndCalib}b. The higher photon-number POVM elements are particularly sensitive to this uncertainty, and show correspondingly large deviations from their ideal values. This highlights the crucial importance of an accurately calibrated probe state source for detector tomography. Our setup has a high calibration uncertainty of 8\%, however, calibration uncertainties of less than 1\% are achievable~\cite{Lunghi2014,Miller2011}. Since this shortcoming is not intrinsic to our detector, in the following analysis we will assume such a 1\% calibration uncertainty, as this allows us to better demonstrate the information that our protocol can provide.


\emph{Characterising photon-number resolution:} The above tomography procedure gives the probability density $p(s | n)$ for a specific outcome $s$ given an $n$-photon input to the detector.  However, in typical experiments, we are actually interested in the complementary probability density $p(n | s)$ that the input contained $n$ photons given that the detector measured outcome $s$. Determining this requires Bayes' theorem $p(n | s) = p(s | n) p(n) / p(s)$ and thus depends on our prior probability $p(n)$ of an $n$-photon input~\footnote{See Supplementary Information: Section V for a detailed discussion of the impact of photon-number prior probabilities.}. 
 
\begin{figure}[htbp]
\begin{center}
\includegraphics[width=8cm]{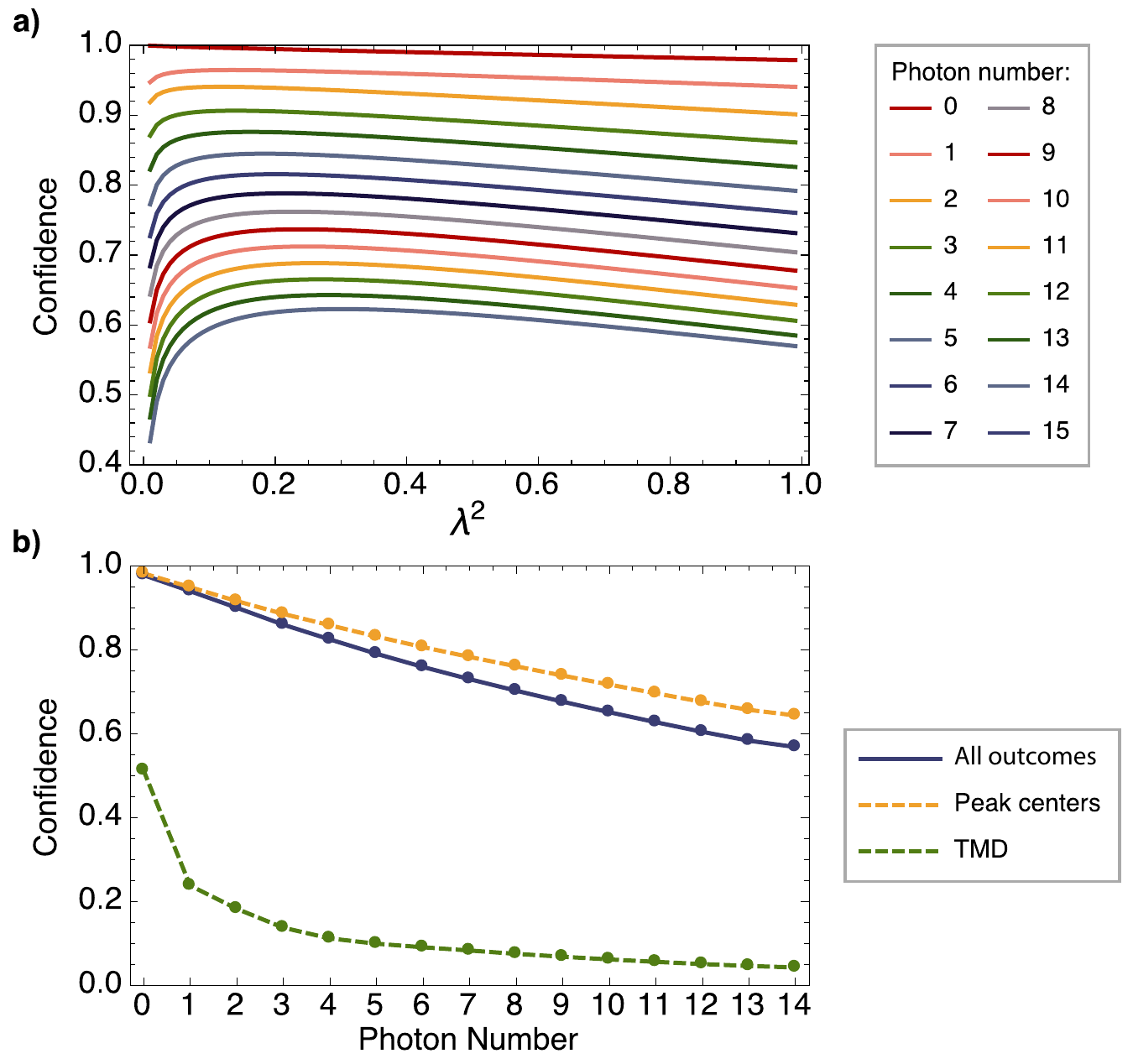}
\caption{a) Calculated confidence $C_n$ for different photon numbers as a function of the thermal prior distribution parameter $\lambda^2$. b) Calculated confidence for our detector given a flat prior, as a function of photon number $n$ (blue). Confidence for outcomes at the centres of the peaks in $p(s | n)$  (dashed yellow). Confidence for a time-multiplexed pseudo-number-resolving detector (dashed green).}
\label{fig:Confidence}
\end{center}
\end{figure}

Closely linked to determining $p(n | s)$ is the problem of finding a quantitative measure of the `photon-number resolution' of the detector. Since $p(n | s)$ only gives information on the confidence with which a specific outcome $s$ can determine the photon-number input, we propose a measure that represents an average of this confidence, weighted by the probability density for $s$ given $n$ input photons,
\begin{align}
C_n &= \int^\infty_{-\infty} p(n | s) p(s | n) \, \mathrm{d}s= \int^\infty_{-\infty} \frac{p(s | n)^2 p(n)}{p(s)}  \, \mathrm{d}s \nonumber \\
&= \int^\infty_{-\infty} \frac{p(s | n)^2 p(n)}{\sum_k p(s | k) p(k)}  \, \mathrm{d}s. \notag
\end{align}
Given an input of $n$ photons, this confidence $C_n$ represents the average probability ascribed to the $n$ photon component of the inferred state $\rho(s) = \sum_n p(n | s) \ket{n}\!\!\bra{n}$. More loosely, it represents the probability that the detector gives the correct photon number. Additionally, $C_n = \int  \bra{n} \rho(s) \ket{n} p(s | n) \mathrm{d}s$, the average squared fidelity between the inferred detected state and an $n$ photon number state $\ket{n}$, weighted by the probability $p(s|n)$. For the detection of a heralding state from a spontaneous parametric down-conversion (SPDC) source~\cite{Ramelow2012}, this will therefore also be the fidelity of the heralded state with $\ket{n}$. Note that the detector does not have information on the specific input photon-number $n$; however, a prior distribution must be specified. This confidence is therefore a function of the distribution chosen. Fig~\ref{fig:Confidence}a shows the confidence for different photon numbers as a function of the SPDC source thermal prior distribution parameter $\lambda^2$, where $p(\ket{n,n} | \, \lambda) = (1 - \lambda^2) \lambda^{2 n}$.

In order to facilitate comparison between different detectors, it may be useful to determine this confidence given a flat prior for the photon number,
\begin{equation}
C_n = \int^\infty_{-\infty} \frac{p(s | n)^2}{\sum_k p(s | k)}  \, \mathrm{d}s. \notag
\end{equation}
This is plotted in Fig.~\ref{fig:Confidence}b. As would be expected, our detector is extremely effective at resolving vacuum and lower photon numbers, while for higher photon numbers, the increasing effect of the detection inefficiency and gradual saturation of the detector leads to a reduced confidence in the outcomes. As an example of the additional information given by our continuous-output analysis, we also plot the confidence for a post-selected case, in which only outcomes at the centres of the peaks in $p(s | n)$ (Fig.~\ref{fig:EMsolutionAndCalib}) are accepted~\footnote{See Supplementary Information: Section VI for a further discussion on the use of post-selection to increase the detector confidence.}. This could be employed to boost the fidelity of the heralded Fock states produced by SPDC sources. In order to demonstrate that this measure is widely applicable to different PNRDs, the confidence for the time-multiplexed pseudo-number-resolving detector with 8 time bins presented in~\cite{Lundeen2008} is also shown.

\section{Acknowledgements} 
This work was supported by the UK Engineering and Physical Sciences Research Council (EPSRC EP/K034480/1) and the European Office of Aerospace Research \& Development (AFOSR EOARD; FA8655-09-1-3020). WSK is supported by EC Marie Curie fellowship (PIEF-GA-2012-331859). AD is supported by the EPSRC (EP/K04057X/1). We thank Alvaro Feito for kindly helping us to obtain the data from his previous paper and Tim Bartley for his help in installing the TES detectors. Contribution of NIST, an agency of the U.S. Government, not subject to copyright.

\bibliography{TESpapers}

\section{Principal component analysis}
 
\begin{figure}[htbp]
\begin{center}
\includegraphics[width=8.5cm]{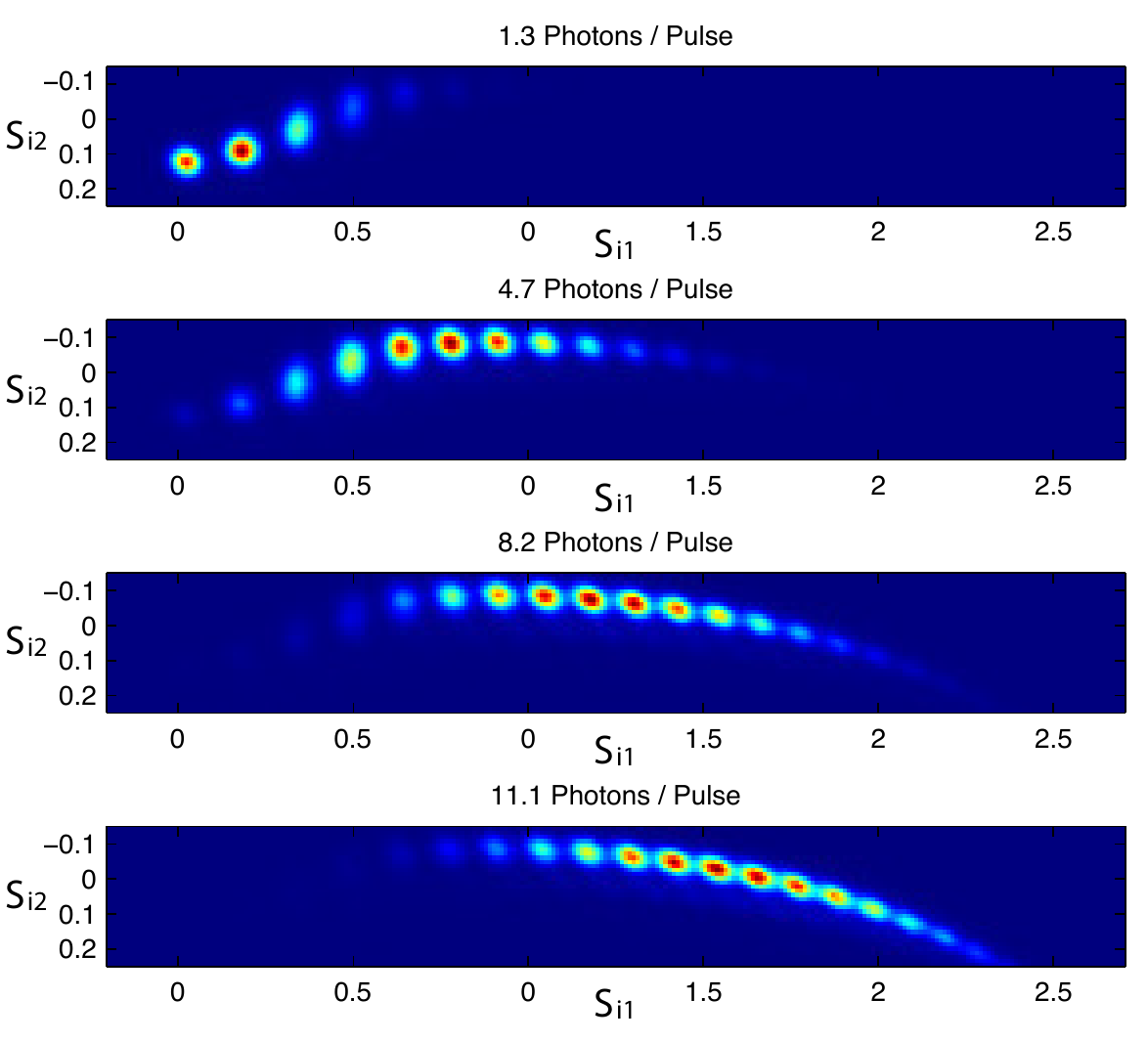}
\caption{Example probability density functions in the space of the first and second principle component scores for coherent state probes with varying average photon numbers.}
\label{fig:exampleSVDPDF}
\end{center}
\end{figure}

The main text of the paper focuses on only the first principal component $w_1(t)$. Although, particularly for lower photon numbers, this component provides most of the distinguishing information available (as measured through the data covariance), it is interesting to note that higher order components can contribute additional information. In Fig.~\ref{fig:exampleSVDPDF} we plot example probability density functions in the space of $s_{i1}$ and $s_{i2}$ for different coherent state probes. Structure along $s_{i2}$ is visible, and could be incorporated into a detector tomography analysis to further distinguish input states.

\section{Gaussian-mixture maximum likelihood estimation}

\begin{figure}[htbp]
\begin{center}
\includegraphics[width=8.5cm]{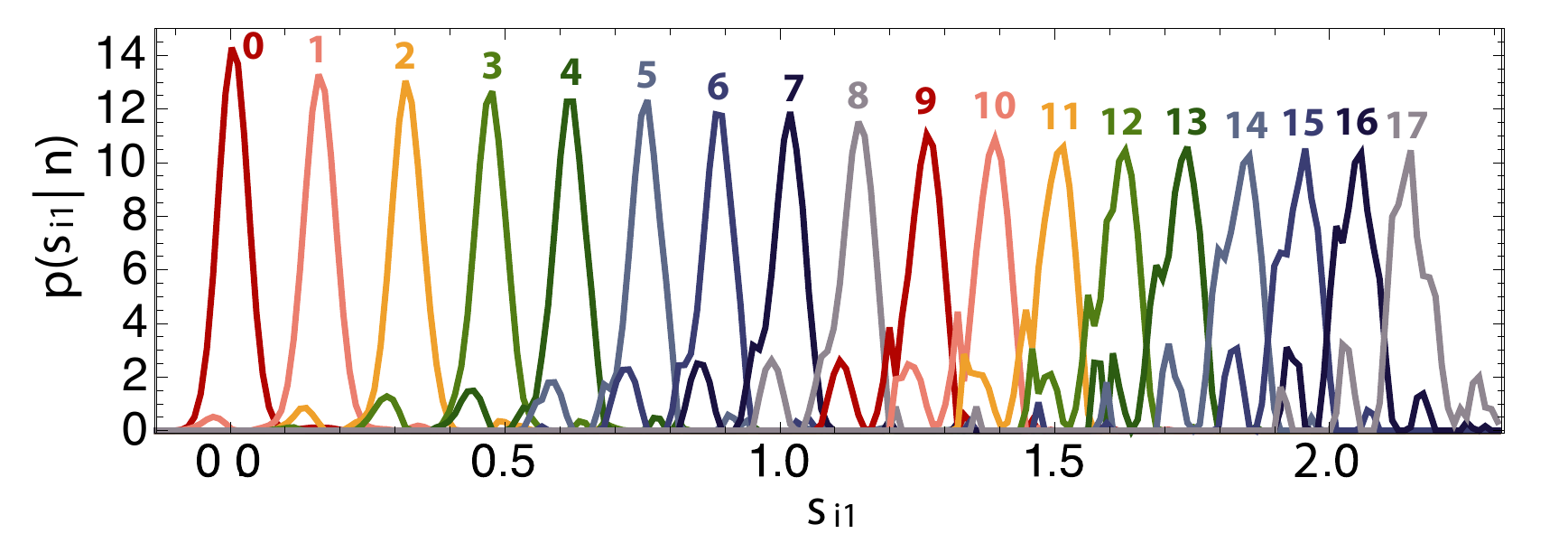}
\caption{Constrained least squares solutions for the Fock state POVM elements $\Pi_{n,s}$ showing the responses to vacuum and up to 17 photons. The solution has unphysical noise features.}
\label{fig:SOCPsolution}
\end{center}
\end{figure}

We found that published methods of performing detector tomography based on constrained least squares techniques~\cite{Brida2012a} did not give satisfactory results, with the resulting POVM clearly showing unphysical noise features (Fig.~\ref{fig:SOCPsolution}). This may be because the reduced overlap between the POVM elements for different photon numbers as compared to previous studies means that regularisation techniques designed to promote this overlap~\cite{Lundeen2008} do not work as effectively. 

As introduced in the main text, we have developed a novel detector tomography routine that is effective for high quantum efficiencies. In this approach, we parameterise the detector response in $s_{i1}$ as a series of overlapping Gaussian distributions. Specifically, we model the response of our detector to a photon number $n$ as composed of a sum of $n+1$ Gaussians, with widths, heights, and positions as free variables. Our approach is readily extendable to higher order components ($s_{i2}, s_{i3} \dots$), however, as in the main text, we choose to focus on $s_{i1}$ here. This model gives the following expression for the POVM coefficients for photon number $n$,
\begin{equation}  
\theta_{n,s} = \sum_j \beta_{n,j} \, \mathcal{N} (s | \mu_{n,j}, \sigma_{n,j}).\label{eqn:modelPOVMelem}
\end{equation}
where $\beta_{n,j}$ is a weighting factor for the Gaussian probability distribution $\mathcal{N} (s | \mu_{n,j}, \sigma_{n,j})$ in the outcome space $s$, with mean $\mu_{n,j}$, and standard deviation $\sigma_{n,j}$.

We imposed the constraint that $\mu_{n,j}$ = $\mu_{n+1,j}$, i.e. that the Gaussians from different photon numbers should be aligned. This is physically motivated by the fact that the detector cannot distinguish between cases where $n$ photons were input and cases where $n+1$ photons were input and one photon was lost. Removing this constraint does not alter the solution significantly, beyond leading to a slight jitter in the location of the peaks for each photon number. However, this jitter complicates the additional analysis that we carry out, particularly with regard to compensating for the uncertainty in the probe state energies (as discussed in Section~\ref{sec:CalibLightSource}). No constraint is placed on $\sigma_{n,j}$.

Substituting equation~(\ref{eqn:modelPOVMelem}) into equation~(4) of the main text, we find that 
\begin{equation}
p(s  | \alpha, \v{\chi}) = \sum_{n,j} F_{\alpha,n}  \beta_{n,j} \mathcal{N} (s | \mu_{n,j}, \sigma_{n,j}),
\end{equation}
where $\v{\chi}$ is used a shorthand to denote the set of all the parameters $\beta_{n,j}, \mu_{n,j}, \sigma_{n,j}$, in order to make the dependence on the model explicit.

This expression gives the posterior probability density for the TES detector producing an outcome $s$ in our model, given an input coherent-state probe $\ket{\alpha}\!\!\bra{\alpha}$. We wish to maximise this posterior probability for the data that we measure. Typically, maximum likelihood estimation~\cite{Bishop2006} is carried out based on a set of observed outcomes $\{ s_{i1} \}$. In this case, the quantity to be maximised is the log-likelihood
\begin{align}
\mathcal{L} &= \log \left( \prod_i p(s_{i1}  | \alpha_i, \v{\chi}) \right)\notag\\
 &= \sum_i \log \left(\, p(s_{i1} | \alpha_i, \v{\chi}) \,\right).\label{eqn:datasetll}
\end{align}

However, due to the large number of data points that we sample, evaluation of this sum becomes impractical. Instead, we used our data set $\{ s_{i1} \}$ to estimate the outcome probability density $q(s | \alpha_k)$ for each $\ket{\alpha_k}\!\!\bra{\alpha_k}$. We can use this distribution to rewrite equation~(\ref{eqn:datasetll}) as 
\begin{equation}
\mathcal{L}  =  \sum_{k} \int N_k\, q(s | \alpha_k) \log \left(\, p(s | \alpha_k, \v{\chi}) \, \right)  \mathrm{d}s,\notag
\end{equation}
where $N_k$ is the total number of samples measured at each value of $\alpha_k$. Since we measured the same number of samples per coherent state value, we will neglect this constant factor that has no impact on the maximum likelihood estimation.

The full expression for the log-likelihood therefore becomes
\begin{align}
\mathcal{L}  &=  \sum_{k} \int  \mathrm{d}s \, q(s | \alpha_k) \dots \notag\\
& \quad \qquad \log \left(\, \sum_{n,j} F_{\alpha_k,n} \, \beta_{n,j} \, \mathcal{N} (s | \mu_{n,j}, \sigma_{n,j}) \,\right).\notag
\end{align}

In order to maximise this log-likelihood, we follow the standard approach~\cite{Bishop2006} of taking derivatives with respect to each parameter in the model. For example, differentiating with respect to $\mu_{n,j}$ gives
\begin{equation}
\pd{\mathcal{L}}{\mu_{n,j}}  = \sum_{k} \int \, q(s | \alpha_k) \, \gamma_{s,k,n,j} \sigma_{n,j} (s - \mu_{n,j}) \,  \mathrm{d}s \notag
\end{equation}
in which we have defined
\begin{equation}
\gamma_{s,k,n,j} =  \frac{F_{\alpha_k,n} \, \beta_{n,j} \, \mathcal{N} (s | \mu_{n,j}, \sigma_{n,j})}{\sum_{n,j} F_{\alpha_k,n} \, \beta_{n,j} \, \mathcal{N} (s | \mu_{n,j}, \sigma_{n,j})}. \notag
\end{equation}

Rearranging leads to the following expression for $\mu_{n,j}$
\begin{equation}
\mu_{n,j} = \frac{1}{N_{n,j}} \sum_{k} \int \, q(s | \alpha_k) \, \gamma_{s,k,n,j} \, s \,  \mathrm{d}s \label{eqn:muSoln}
\end{equation}
where 
\begin{equation}
N_{n,j} =  \sum_{k} \int \, q(s | \alpha_k) \, \gamma_{s,k,n,j} \, \mathrm{d}s. \notag
\end{equation}

Similarly we find that 
\begin{align}
\sigma_{n,j} &= \frac{1}{N_{n,j}} \int \, q(s | \alpha_k) \, \gamma_{s,k,n,j} \, (s - \mu_{n,j})^2 \,  \mathrm{d}s \label{eqn:sigSoln}
\end{align}
and 
\begin{align}
\beta_{n,j} &= \frac{N_{n,j}}{N_n} \text{, where } N_n = \sum_j N_{n,j} \label{eqn:betaSoln}.
\end{align}

Note that these expressions for the parameters are dependent on $\gamma_{s,k,n,j}$, and therefore do not form a closed-form solution. This means that the optimal solution cannot be found analytically. However, it can be shown that a simple routine consisting of the repeated application of two steps will converge to a solution~\cite{Bishop2006}. In the first step, the current values of the parameters are used to calculate $\gamma_{s,k,n,j}$. This is then used in the second step to re-estimate the optimal values of the parameters using equations~(\ref{eqn:muSoln}), (\ref{eqn:sigSoln}) \& (\ref{eqn:betaSoln}).

\section{Calibrated light source}
\label{sec:CalibLightSource}
\begin{figure}[htbp]
\begin{center}
\includegraphics[width=8.5cm]{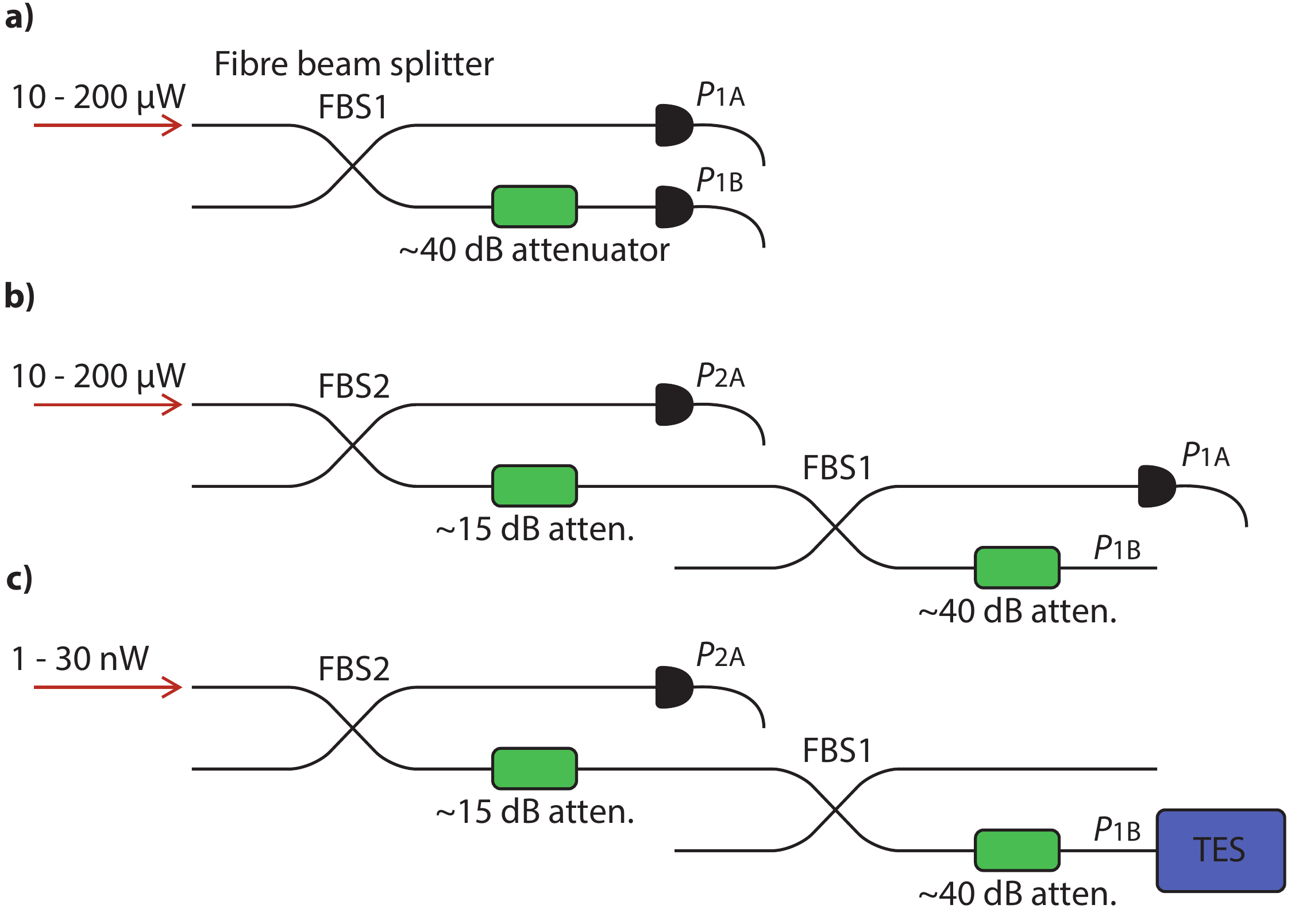}
\caption{Calibrated light source. Because the dynamic range of our power meter is insufficient to span the attenuation required to reduce the coherent-state energy to the single-photon level, we perform the calibration in two steps at the expense of increased error. We use the laser diode running in CW mode to calibrate the attenuators, since the power meter is most accurate in this mode.  a)  We first take a series of measurement of the output powers $P_{1A}$ and $P_{1B}$ for a range of input powers to the fibre beam splitter. This lets us calibrate the output power in port $1B$ if we know the power in port $1A$. b) We connect a second fibre beam splitter to the first one before we calibrate it to make sure the effect of the FC/FC connection is properly accounted for. We then make a series of measurements of $P_{2A}$ and $P_{1A}$ for a range of input powers. Concatenating these results we now know the output power in port $1B$ given the recorded power in $2A$. c) For the detector characterisation we switch the laser to pulsed operation and attenuate the input light to the nanowatt level.
}
\label{fig:calibratedLightSource}
\end{center}
\end{figure}

We built a calibrated coherent state source based on a Newport 918D-IG-OD3R power meter, which provides a specified calibration accuracy of 2\% of absolute power and a linearity of better than 0.5\%. This power meter was used to calibrate a series of fixed attenuators to reduce the output from a pulsed laser to the single-photon level with a known mean-photon number per pulse~\cite{Miller2011}.

Our method uses a fibre beam splitter with a fixed fibre attenuator connected on one of the output ports, as shown in Fig.~\ref{fig:calibratedLightSource}a. As long as this attenuation is well within the linear dynamic range of the power meter, we can obtain a calibration curve for the combined splitter-attenuator device relating the power measured at $P_{1A}$ to the power at $P_{1B}$. In our case, the attenuation required to reach the single photon level is much greater than the dynamic range of the power meter. This forces us to use a second, calibrated splitter-attenuator device in series with the first Fig.~\ref{fig:calibratedLightSource}b. A weighted total least-squares algorithm~\cite{Krystek2007} was used to find the total attenuation taking account of the absolute power errors in both variables. The total attenuation is given by the product of the two attenuators, but the errors in the measurements add linearly since they are not independent. Thus our final calibrated attenuation is found to be
\begin{equation}
\eta_{\mathrm{att}} = (2.10 \pm 0.16) \times 10^{-6}, \notag
\end{equation}
which relates the power measured at the monitor port $2A$ to the power at port $2B$ (Fig~\ref{fig:calibratedLightSource}c). A variable attenuator is used to set the input power level before the calibrated attenuator so that we can probe our detector with a variety of coherent state amplitudes. We monitor the input power to the attenuator using port $2A$ and calculate the average photon number per pulse in port $1B$ which is coupled to the TES.  The value of $\eta_\mathrm{att}$ also includes a correction to account for the Fresnel reflection from the unterminated fibre when plugged into the monitor power meter, which leads us to underestimate the total power that will be input when this fibre is instead directly coupled to the fibre leading to the TES. Fibre specifications give this loss at about 3.3\% but there is a 1\% uncertainty in this figure~\cite{DeCusatis2013}.

As we discuss in the Methods section of the main text, the POVM element coefficient $\theta_{n}(s)$ gives the probability density $p(s | n)$ that we will measure outcome $s$ given $n$ input photons. This probability is actually $p(s | n, \eta_{\mathrm{att}})$ since $\eta_{\mathrm{att}}$ is a variable in our tomography calculations. Our uncertainty in  $\eta_{\mathrm{att}}$  must therefore be accounted for. Based on our error analysis (and assuming normally distributed errors), we can estimate the probability density $p(\eta_{\mathrm{att}})$ for $\eta_{\mathrm{att}}$. Additionally, we can calculate $p(s | n, \eta_{\mathrm{att}})$ for different $\eta_{\mathrm{att}}$. Combining these, we can incorporate this statistical uncertainty into our POVM using\begin{equation}
p(s | n) = \int p(s | n, \eta_{\mathrm{att}}) p(\eta_{\mathrm{att}}) \mathrm{d}\eta_{\mathrm{att}}. \notag
\end{equation}
The results of this analysis are shown in Fig.~2b of the main text.

\section{Estimating the system detection efficiency}
 \begin{figure*}[htbp]
\begin{center}
\includegraphics[width=12.5cm]{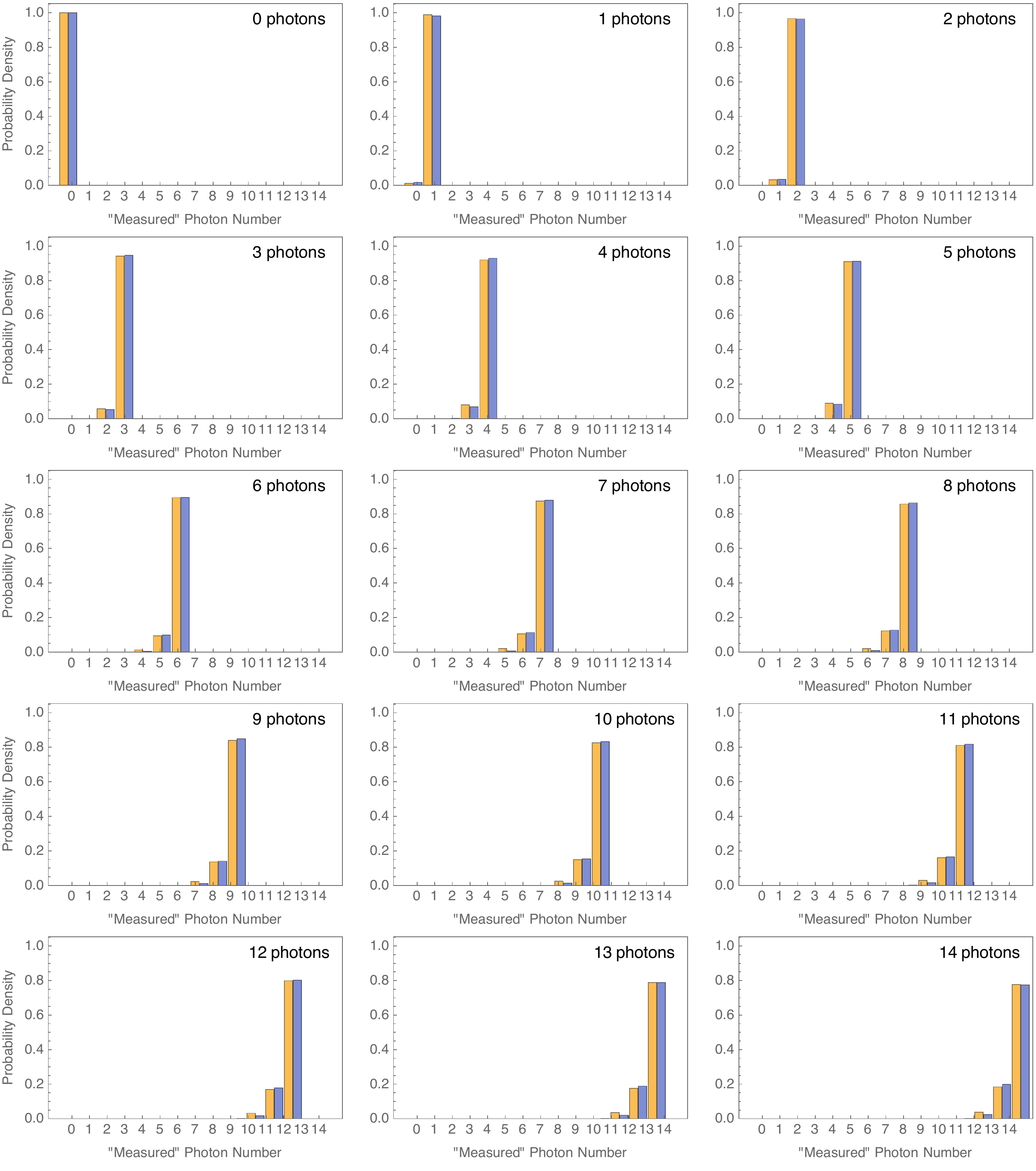}
\caption{Estimating the system detection efficiency. We can assume that the different Gaussian elements that make up the detector response to a given photon number are due to losses between the input and the detection, leading to a reduced number of photons actually being detected. This loss therefore leads to a binomial distribution of Gaussian response elements when an initial Fock state is input to the detector. The relative weights of each of these Gaussian elements, as inferred from our maximum likelihood estimation protocol, are plotted here in gold. From this data, we can carry out a numerical routine to determine the detection efficiency most consistent with our data. The predicted photon (and therefore Gaussian element) distribution resulting from this estimated detection efficiency is shown in blue for comparison.}
\label{fig:binomial}
\end{center}
\end{figure*}

The POVM elements that we calculate using detector tomography completely characterise the detector response. Model-free detector tomography is obtained by treating the detector as a black box, and so in principle does not contain information on the system detection efficiency, i.e. the loss that occurs between the input and the detector. 

However, the less general, but physically motivated model-based detector tomography approach that we have adopted can allow us to make an estimate of this efficiency. As noted above, we assume that the response of the detector to each photon number is composed of several Gaussian elements. We can make the further assumption that these different Gaussian elements occur due to the action of loss on an initial Fock state, leading to a statistical mixture of photon numbers at the detector. Therefore the heights of these elements should follow a binomial distribution within each Fock state POVM element. For a given system detection efficiency, it is then possible to calculate the expected height of these Gaussian elements and compare them to the actual tomography output. We used a numerical routine to find the loss level that minimised the L2 norm between this predicted output and the tomography data. 

This analysis suggests that our system detection efficiency is $0.98 \, ({+0.02}/{-0.08})$. The asymmetric uncertainty arises as the efficiency is upper bounded at 1.0. Additionally, we find a strong agreement between the predicted photon number distribution and the tomography data, as shown in Fig.~\ref{fig:binomial}, suggesting that our initial assumption is correct. 
 
 \section{Impact of photon-number prior probabilities}
 
As we discuss in the main text, detector tomography gives the probability $p(s | n)$ that a specific outcome $s$ will occur given an $n$-photon input to the detector.  However, in typical experiments, we are actually interested in the complementary probability $p(n | s)$ that the input contained $n$ photons given that the detector measured outcome $s$ (as calculated from the detector signal $v(t)$ by $s = \int v(t) w_1(t) \mathrm{d}t$). 

Calculating this requires Bayes' theorem $p(n | s) = p(s | n) p(n) / p(s)$ and thus depends on our prior probability $p(n)$ of an $n$-photon input. Here, as an example we consider two distinct priors which might arise in applications. First, we consider a Poisson distribution $p(n | \alpha) = e^{- \abs{\alpha}^2} \abs{\alpha}^{2 n} / n!$ which would result from a coherent state input. We also consider a thermal distribution $ p(n | \lambda) = (1 - \lambda^2) \lambda^{2 n} $ which describes a thermal state input and, importantly, the single-mode marginal statistics of a spontaneous parametric down-conversion source. If one mode of such a source is sent to a detector, $p(n | s, \lambda)$ represents the statistical mixture of photon numbers onto which the other mode is projected. Such information is extremely important for quantum information and metrology applications.

\begin{figure}[htbp]
\begin{center}
\includegraphics[width=8.0cm]{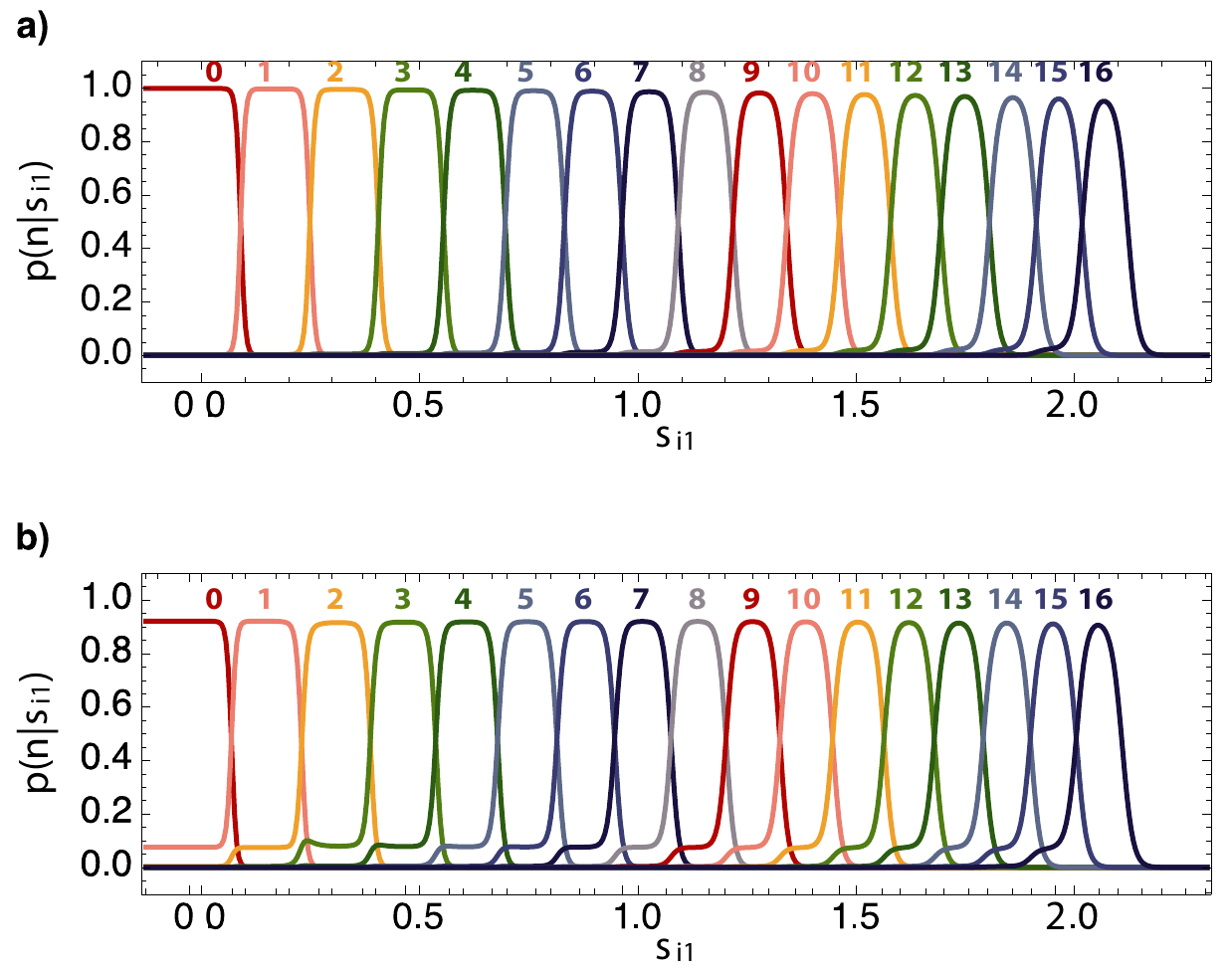}
\caption{Example distributions $p(n | s_{i1})$ from our tomography data, which give the probability that the input contained $n$ photons given that the detector measured outcome $s$. The effect of the prior input photon number probabilities can be seen in the difference between a) a thermal distribution with $\lambda^2 = 0.1,$ and b) a Poisson distribution with $\abs{\alpha}^2 = 5$.}
\label{fig:BayesCalculations}
\end{center}
\end{figure}

Two example probability distributions $p(n | s, \alpha)$ and $p(n | s, \lambda)$ are plotted in Figs.~\ref{fig:BayesCalculations}~a~\&~b. As can be seen, the two priors lead to significant differences in the distributions. For the thermal distribution, the thermal prior suppresses the overlap between the outcomes associated with neighbouring photon numbers. This is because, for small $\lambda$, $n+1$ input photons will occur much less frequently than $n$ photons. Therefore the predominant overlap contribution, due to an $n+1$-photon input being detected in the space of outcomes most associated with $n$ input photons, occurs correspondingly less frequently than genuine $n$-photon inputs.

The Poissonian prior plotted in Fig.~\ref{fig:BayesCalculations}b has the opposite effect as the thermal prior, since in this case an input of $n+1$ photons is more probable than an input of $n$ photons, and therefore the overlap is promoted. It should be noted that in both cases, due to the truncation of our detector tomography at 17 input photons, the distributions $p(n | s)$ become inaccurate in regions in which significant contributions would be expected from photon numbers greater than this. In practice, this simply translates to an operational requirement that detector tomography must be extended to include all photon numbers that are expected to contribute in any given experiment.

\section{Post-selecting outcomes to improve confidence}

For certain applications~\cite{Datta2011}, it is important to maximise the fidelity of the inferred detected state with a photon number state ($C_n$). In these cases, the fidelity can be improved using post-selection strategies in which only a subset of outcomes are accepted. This is possible to explore using our detector tomography data since our treatment has explicitly avoided any binning of outcomes. 

\begin{figure}[htbp]
\begin{center}
\includegraphics[width=8.0cm]{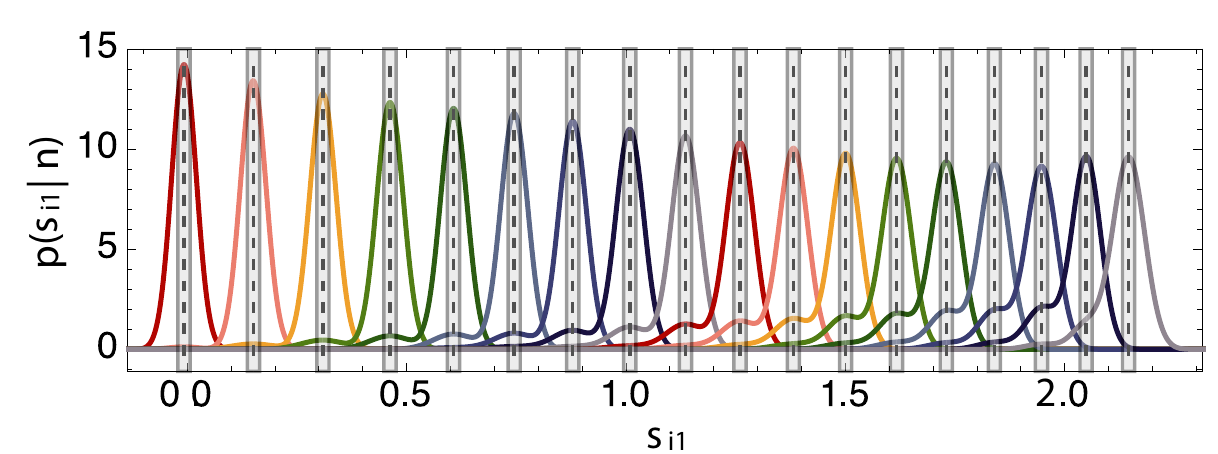}
\caption{Post-selecting on outcomes within windows centred on the peak maxima can be employed to boost the confidence of detected photon states.}
\label{fig:PostSelect}
\end{center}
\end{figure}

One strategy is to only consider outcomes within windows centred on the peak maxima (Fig.~\ref{fig:PostSelect}). As would be expected, the highest confidence is obtained in the limit of the window width tending to zero, in which case the number of accepted outcomes would also tend to zero. This limit therefore upper bounds the performance of this strategy, and is plotted in Fig~3b of the main text. For our detector, the increase in confidence as compared to using the full space of outcomes is comparatively modest, since the overlap between different photon number POVM elements is dominated by the detection efficiency. However, as the detection efficiency of detectors improves, the intrinsic overlap between neighbouring Gaussian peaks is expected to become increasingly important. In this case, this post-selection strategy should become more effective.

\end{document}